# How Social Media Big Data Can Improve Suicide Prevention

Anastasia Peshkovskaya[1, 2], Yu-Tao Xiang[3, 4]

[1] Tomsk State University, Tomsk, Russia

[2] Mental Health Research Institute, Tomsk National Research Medical Center, Russian Academy of Sciences, Tomsk, Russia

[3] Unit of Psychiatry, Department of Public Health and Medicinal Administration, Institute of Translational Medicine, Faculty of Health Sciences, University of Macau, Macao SAR, China

[4] Centre for Cognitive and Brain Sciences, University of Macau, Macao SAR, China

Funding statement

The linguistic toolkit was developed within research project No.19-78-10122 under RSF Grant.

Acknowledgment

AP would like to thank members of the Center of Applied Big Data Analysis at Tomsk State University who were responsible for the linguistic toolkit efficacy in social media's data search conditions.

# How Social Media Big Data Can Improve Suicide Prevention

**Abstract**

In the light of increasing clues on social media impact on self-harm and suicide risks, there is still no evidence on who are and how factually engaged in suicide-related online behaviors. This study reports new findings of high-performance supercomputing investigation of publicly accessible big data sourced from one of the world-largest social networking site. Three-month supercomputer searching resulted in 570,156 young adult users who consumed suicide-related information on social media. Most of them were 21-24 year olds with higher share of females (58.51%) of predominantly younger age. Every eight user was alarmingly engrossed with up to 15 suicide-related online groups. Evidently, suicide groups on social media are highly underrated public health issue that might weaken the prevention efforts. Suicide prevention strategies that target social media users must be implemented extensively. While major gap in functional understanding of technologies' relevance for use in public mental health still exists, current findings act for better understanding digital technologies' utility for translational advance and offer relevant evidence-based framework for improving suicide prevention in general population.

**Main**

Suicide costs are high, as 800 000 people die by suicide annually (WHO, 2019). Burden of pandemic-linked suicide might be a critical public health issue in the nearest future, as COVID-19 pandemic impact on mental health unfolds. Over the last years, increasing evidence revealed the mediating role of suicide-related social media use and suicidal risks (Janiri et al., 2020; Luxton, June, Fairall, 2012). Social media suicide groups that romanticize suicidal deaths and their means can influence suicide-related ideation and possibly lead to an increased rate of self-injuries and suicides (Liu et al., 2020; Peshkovskaya et al., 2020; Pourmand et al., 2019). However, little is known about suicide-related social media engagement and who participates in. For the most part, highly vulnerable to suicide-related media content can be youth, a group with dominance on sites such as Instagram, Facebook, TikTok, and others (Peshkovskaya et al., 2021).

In this work, we provide results on three-month data collection on suicide-related groups on one of the world-largest social networking site. We aimed our study at investigating suicide group followers, particularly their publicly disclosed basic population characteristics. During January-March 2020, we collected open and publicly accessible data of the largest European social networking site VK (vk.com) with more than 100 million active users. As big data analysis requires significant resources and computing power, supercomputer with 6.2 thousand computing cores, high-speed Internet connection, and a high-performance cluster for collecting, storing and processing data was used. Big data search algorithm employed linguistic toolkit based on vocabulary of eight words and word combinations, such as "suicide", "suicide methods", "best suicide methods", "to kill yourself", "wanna die", "better to die". The study was approved by University Research Consortium Ethics Committee and conducted according to the Federal Law on Data Protection, the Declaration of Helsinki, and the Human Rights Act. Study findings reported in accordance with JARS-Quant Standards on new data collections.

As a result of a three-month data collection, we found 24 suicide-related online groups with 570,156 followers in total. While the lowest group followers' number started from 1,267, the largest group consisted of 209,478 of those. Larger share of suicide-related groups' audience was females, who approximated 58.51% of the followers aged 21.25 years in average. Males were of 24.33 years of age and amounted 41.49% of the whole sample (Figure 1).

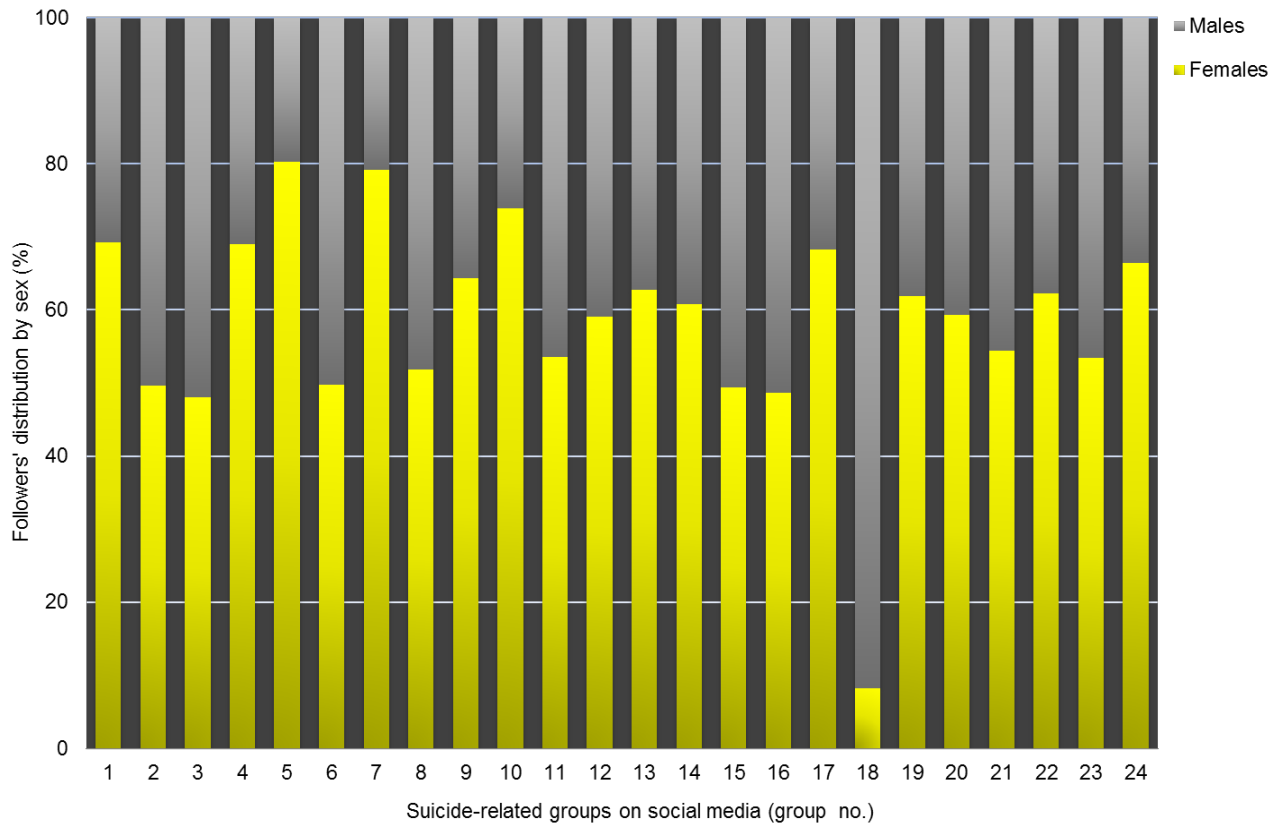

**Figure 1**. Followers' distribution by sex (%). Open and publicly accessible data on basic demographics.

Employing profile-matching algorithm, we validated 70,881 social media users who followed two or more suicide-related groups. Share of those reached 12.43% of total follower's number and upward previously founded male-female ratio toward greater female share of 67.75% (48,024 female users in total). Amount of males lowered to 35.25% (22,857 users) respectively (Table 1). Basically, every eight suicide group follower appeared to be exceptionally interested in suicide-related information and followed at least two or utterly fifteen suicide-related groups on social media.

**Table 1**. Social media users who followed two or more suicide-related groups: Descriptive statistics. Open and publicly accessible data on basic demographics.

| Amount of followed suicide-related groups per user | Amount of users | Sex | Mean Age |
|---|---|---|---|

|   |   | Females | Males |   |
|---|---|---|---|---|
| 2 | 49,929 | 34,288 | 15,641 | 21.6 |
| 3 | 12,941 | 8,656 | 4,285 | 21.7 |
| 4 | 4,392 | 2,850 | 1,542 | 21.2 |
| 5 | 1,810 | 1,149 | 661 | 20.7 |
| 6 | 851 | 517 | 334 | 21.8 |
| 7 | 433 | 273 | 160 | 21.3 |
| 8 | 242 | 137 | 105 | 21.7 |
| 9 | 124 | 66 | 58 | 26.1 |
| 10 | 78 | 41 | 37 | 18.4 |
| 11 | 42 | 26 | 16 | 23.9 |
| 12 | 24 | 11 | 13 | 24.7 |
| 13 | 9 | 6 | 3 | 19.5 |
| 14 | 4 | 2 | 2 | 34.5 |
| 15 | 2 | 2 | 0 | 22.0 |
| Mean 8.5 | Total 70,881 | Total 48,024 | Total 22,857 | Overall Mean 21.6 |

**Prevention programs must target internet users**

Overall, more than half million young adults consuming suicide-related information were found in three-month supercomputer search on social media. Their number might be overwhelmingly higher as we continue the data collection. Evidently, suicide groups on social media are highly underrated public health issue that might weaken the prevention efforts. Our findings clearly indicate that prevention programs must address internet users. While telehealth technologies including online social therapy are already employed for people with active suicidal ideation (Bailey et al., 2021), covering larger young adult population on social media might be beneficial as suicide is the fourth leading cause of death in 15-29 year-olds globally (WHO, 2021). Considering widespread concern on COVID-19 impact on mental health (Cai et al., 2023; Peshkovskaya, 2021), and pandemic-related increase of internet use (Pew Research Center, 2021), prevention strategies that address social media users will forge ahead existing interventions and must be planned extensively.

Likewise, interaction with the media for responsible reporting of suicide-related information is inevitable. Based on the reported age-related findings, we suggest that age-based content rating may also be considered or even raised in regard to suicide-related information on media to attenuate any possible risk for youth of being affected by any information on suicide or its methods.

**The insights for vulnerability understanding**

We consider that 70,881 young adults who followed up to fifteen online suicide groups showed exceedingly red flagging social media use behavior. In view of increasing evidence on social media impact on self-harming behaviors and suicide risk (Liu et al, 2020; Voevodin, et al., 2020), it is strongly recommended to develop early identification of anyone who is affected by massive suicide-related information online.

**Big Data-based framework for improving public health interventions**

Notwithstanding the extraordinary rise of digital health amid COVID-19 crisis, including health monitoring apps, video-based telehealth, and social networking platforms and messengers to meet patients' needs for affordable, convenient and readily-accessible mental health services (Nature Medicine, 2021; McKendry et l., 2020; Budd et al., 2020), many are still unaware of how technological innovations and large amount of data they generated can be applicable and safely utilized. Obviously, there is a strong need in functional understanding of digital technologies applications which are relevant to clinical and research use in psychiatry (Lundin, Menkes, 2021) and mental illness prevention (Torous, et al., 2018). Since 2016, when European Commission developed recommendations for digital technologies use in prevention and health care in the European Union (European Commission, 2016), studies have been emerged to identify big data applications for the individual and the whole population health. While massive data in health care is already being generated and ready for use from various sources, social media is undoubtedly one of those evolving data sources which are offer large opportunities for research in prevalence and prevention. This study and some of its early results presented and awarded at the ECNP Congress 2021 (Peshkovskaya A, Matsuta V., 2021) act as functional example of how big data sourced from social media is employed accordingly with privacy protection regulations for better understanding suicide-related behavior. Described results offer the evidence-based framework for planning targeted interventions and improving suicide prevention in general population.

Data Availability Statement

Requests for raw and analyzed data may be directed to Anastasia Peshkovskaya and will be reviewed by the board of the Center of Applied Big Data Analysis to determine whether the request is subject to any intellectual property or confidentiality and data protection obligations. Data that can be shared will be released via a material transfer agreement. Source data for figure 1 are provided with the paper.